# Expanding to Arbitrary Study Designs: ANOVA to Estimate Limits of Agreement for MRMC Studies

Si Wen[a]* and Brandon D. Gallas[a]

[a]*Division of Imaging, Diagnostics, and Software Reliability, Office of Science and Engineering Laboratories, Center for Devices and Radiological Health, U.S. FDA, Silver Spring, USA*

* Correspondence to: Si Wen. Email: si.wen@fda.hhs.gov

# Expanding to Arbitrary Study Designs: ANOVA to Estimate Limits of Agreement for MRMC Studies

A multi-reader multi-case (MRMC) analysis is applied to account for both reader and case variability when evaluating the clinical performance of a medical imaging device or reader performance under different reading modalities. For a clinical task that measures a quantitative biomarker an agreement analysis, such as limits of agreement (LOA), can be used. In this work, we decompose the total variation in the data using a three-way mixed effect ANOVA model to estimate the MRMC variance of individual differences and the LOA between different reading modalities. There are rules for writing down the expectation of the mean squares in terms of the variance components for fully-crossed data, i.e. data where all the readers read all the cases in all modalities being studied. Sometimes the annotation task is labor-intensive and time-consuming or distributed across sites, so that a fully-crossed study is not practical. In this work, we focus on estimating the MRMC variance in the within- and between-readers and within- and between-modalities LOA for an arbitrary study design. Simulation studies were conducted to validate the LOA variance estimates. The method was also applied to a dataset to compare pathologist performance for assessing the density of stromal tumor infiltrating lymphocytes on different platforms.

## 1.Introduction

A multi-reader multi-case (MRMC) study is usually applied to evaluate whether a medical imaging device can improve the clinical performance of the image readers.(Wagner et al. 2007; Gallas et al. 2012; Obuchowski and Bullen 2022) In the study, qualitative or quantitative assessments are collected and compared from multiple readers (radiologists/pathologists) reviewing multiple cases (images) under different reading modalities. For example, suppose that there is an AI/ML algorithm enabled medical device supporting the pathologists to evaluate the density of stromal tumor infiltrating lymphocytes (sTILs) on digital slides. The density of sTILs is prognostic for survival in breast cancers and is visually assessed on routine hematoxylin and eosin (H&E)-stained slides.(Kos et al. 2020) To validate the performance of the device in the hands of the readers we collect estimates of the density of sTILs from pathologists with and without the assistance of the device. Then we compare the closeness or agreement between the quantitative measurements, sTIL scores, from different modalities or readers. We refer to such studies as agreement studies.

Our previous work(Wen and Gallas 2022) generalized the limits of agreement (LOA) method(Bland and Altman 1986, 1999), a widely used agreement analysis, to MRMC analysis that accounts for reader and case variability for fully-crossed study designs. In a fully-crossed study, all the readers provide measurements for all the cases for both modalities as shown in Figure 1.e. In some studies, the annotation tasks are labor-intensive and time-consuming or distributed across sites, so that a fully-crossed study is not practical. In this work, we focus on MRMC studies that are not fully crossed, and we generalize the MRMC analysis to treat arbitrary study designs. Some examples are described in Figure 1.a-Figure 1.d.

An ANOVA model is commonly used to analyze MRMC data,(Beiden et al. 2000; Dorfman et al. 1992; Gallas et al. 2009; Obuchowski 1995) as the data are likely correlated when the readers evaluate the same set of cases and each case is reviewed by multiple readers. A three-way mixed effect ANOVA is used to estimate different types of LOA for a fully-crossed MRMC study.(Wen and Gallas 2022) In ANOVA, the data generated from a fully-crossed MRMC study are called balanced data, as each reader or case has the same number of readings. There are rules for calculating the sum of squares (SS) and expected mean squares (MS) when the data are balanced.(Montgomery 2012) The expected mean squares link the variance components for the random effects in the ANOVA model to the SS computed from the data. Then, the variance components can be estimated from the data and used to estimate the MRMC variance in LOA. However, in an arbitrary MRMC study, the data are unbalanced. To determine the relationship between expected MS and the variance components in the ANOVA model, we first need to compute the SS. There are three different classical approaches for computing SS for unbalanced data, commonly called Type I, II and III SS. This notation seems to have been introduced into the statistics world from the SAS package(Langsrud 2003) but is now widespread.

As an example, Table 1 shows Type I, II, III SS structures for each source of variation in a two-way random effect ANOVA model with no replicates. There are two random effects, reader and case, and a random error in the ANOVA model. $RSS()$ denotes the residual sum of squares (RSS) for a multi-variable linear regression. More details about these models and the calculation of their RSS will be discussed in section 2.1. The Type I SS, also called sequential SS, computes the SS for each effect as the reduction in RSS obtained by adding each effect to the regression model. Each effect is considered one at a time as it is added to the model. Thus, the order of the effects matters when estimating the variance components. In Table 1, we present the two possible orders: reader effect first and case effect first. One nice property of the Type I SS is that the SS for all the effects add up to the total SS, $RSS(1)$, as shown in the last row of the first two columns of Table 1. This holds for all the types of SS for balanced data but is not generally true for other types of SS when the data are unbalanced. The Type II SS is called hierarchical or partially sequential SS. Each row is the reduction in RSS due to adding the effect to the model after all other effects has been added to the model, except those that contain the new effect. Unlike Type I SS, Type II SS is invariant to the order in which effects are entered into the model. The Type III SS is called marginal or orthogonal SS, as the SS is the reduction in RSS when the effect is entered last into the model. This means that the interaction term could be added to the model when the main effects it contains are not included.(Montgomery 2012) We have found that this might cause issues when estimating the variance components using Type III SS.

Yates(Yates 1934) described all three methods for two-way tables. Unlike the study designs shown in Figure 1, the two-way tables discussed by Yates can have different numbers of observations/replicates for each reader-case combination. Thus, the study design for a single modality can be regarded as a special case of the two-way tables. However, Yates did not discuss how to derive the expectation of different types of SS and how to estimate the variance components when the effects are random. There are several works in the literature that discuss estimating variance components by using ANOVA models for unbalanced data.(Montgomery 2012; Henderson 1953; Searle et al. 2009; Eze and Nwankwo 2016; Okereke et al. 2020) To the best of our knowledge, there is no literature illustrating how to estimate variance components using all Type I, II and III SS methods and comparing the results. Also, most of the works in the literature focus on one-way or two-way ANOVA models. Okereke et al.(Okereke et al. 2020) presented how to estimate variance component in a three-way mixed-effects model with only one random effect. To analyze MRMC data, we usually set both readers and cases as random effects and the modality is a fixed effect, so that the result of the analysis can be generalized to the group of readers and the group of cases that follows the same distribution. Thus, we need to estimate variance components for a three-way mixed effect ANOVA model with two random effects.

In this article, we will illustrate how to estimate the variance components for the random effects in a two-way random effects ANOVA model and a three-way mixed effects ANOVA model for unbalanced data. From the estimated variance components, we will construct different types of LOA for an MRMC agreement study with an arbitrary study design. The LOA differ depending on whether the comparisons are within- or between-readers and within- or between-modalities. Section 2 will start with the annotations and definitions of different types of comparisons. We simulated MRMC studies with different parameter settings to validate and compare the LOA variance estimates from Type I, II and III SS methods. The methods were also applied to a dataset to compare pathologist performance for assessing the density of sTIL on different platforms. The results for the simulation study and the real data example will be illustrated in section 3. Finally, we will conclude the paper with some discussions in section 4.

## 2. Method

Suppose $X_{ijk}$ denotes the measurement for case $k$ ($k = 1, \dots, K$) from reader $j$ ($j = 1, \dots, J$) under modality $i$ ($i = 1,2$) in an MRMC study comparing two different modalities, where $i = 1$ and $i = 2$ indicate test modality and reference modality respectively. In our previous work,(Wen and Gallas 2022) we introduced four types of MRMC LOA for four types of differences possible in a fully-crossed MRMC study, as shown in Table 2.

If we focus on the within-reader between- modality (WRBM) comparison, we first calculate difference score between two modalities for each reader and case combination. Then, there are only two effects for the difference score: reader and case. Since there are no replicate readings from a reader for a case under each modality, the number of observations for each reader-case combination is just one (when the reader reads the case in both modalities) or zero (when the reading for at least one modality is missed). As we mentioned earlier, it is a special situation of the two-way table discussed by Yates(Yates 1934) and we will apply the methods proposed by Yates to derive the related RSS in section 2.1. There we show how the expectation of the RSS can be linked to the variance components in explicit form. Then, the WRBM limits of agreement can be built given those variance components.

When considering the between-reader between-modality (BRBM) comparison, instead of a two-way table, we need a three-way table to illustrate the design of the study with three effects: reader, case and modality. The three types of SS cannot be expressed explicitly anymore. We will discuss how to estimate the SS from residual sums of squares of linear models and the derivation of its expectation in section 2.2. The theoretical derivations are followed by a simulation study and a real data example described in section 2.3 and section 2.4, respectively.

### *2.1 Using two-way mixed effect ANOVA to estimate the variance of WRBM*

To estimate the variance of the WRBM difference, we build a two-way random effect ANOVA model for the difference $D_{jj,k}^{12}$:

$$D_{jj,k}^{12} = \mu + R_j + C_k + \varepsilon_{jk} \tag{1}$$

where $\mu$ denotes the grand mean, which indicates the average difference in quantitative values between the two modalities, $R_j \sim N(0, \sigma_{2w,R}^2)$ and $C_k \sim N(0, \sigma_{2w,C}^2)$ are random effects for the reader and case respectively, and $\varepsilon_{jk} \sim N(0, \sigma_{2w,\varepsilon}^2)$ represent both the interaction term and random error, as there are no replicates. The $2w$ in the subscript of the variance components denotes they are for the 2-way ANOVA model. From this model, the variance of WRBM differences can be derived as:

$$V_{WR}^{12} = Var\left[D_{jj,k}^{12}\right] = \sigma_{2w,R}^2 + \sigma_{2w,C}^2 + \sigma_{2w,\varepsilon}^2 \tag{2}$$

Table 1 shows the SS structures for each source of variation based on the definition of Type I, II, and III SS. There are four linear models that depend on whether the reader or case effect are involved. Table 3 presents the RSS and their expectation for each of the four models.

We use $r_j$ and $c_k$ in these models to indicate they are fixed effects. The $n_{jk}$ denotes whether the reader $j$ reads case $k$ for both modalities; if $n_{jk} = 1$, we use $d_{jk}$ to denote the realization of $D_{jj,k}^{12}$, otherwise $n_{jk} = 0$, and that difference is missing and left out. Therefore, the RSS for each of the models is the sum of the squared differences between the observed difference score and its prediction given the model.

We listed the RSS for the first three models in Table 3. There is no explicit form for $RSS(R, C)$ (full model), but it can be solved indirectly as follows. First, to estimate $\mu$, $r_j$ and $c_k$ in that model, we need to solve the following system of equations:

$$\begin{gathered} \sum_k d_{jk}\, n_{jk} = \left(\hat{\mu} + r_j\right) \sum_k n_{jk} + \sum_k \hat{c}_k n_{jk}, \forall j \\ \sum_j y_{jk} n_{jk} = \sum_j \left(\hat{\mu} + \hat{r}_j + \hat{c}_k\right) n_{jk}, \forall k \\ \sum_j \hat{r}_j = 0 \end{gathered} \tag{3}$$

This system has $J + K + 1$ equations: one for every reader, one for every case, and a constraint to ensure a unique solution. After solving for $\hat{\mu}$, $\hat{r}_j$, and $\hat{c}_k$, we can calculate $RSS(R, C) = \sum_j \sum_k \left(d_{jk} - \hat{\mu} - \hat{r}_j - \hat{c}_k\right)^2 n_{jk}$. The expectation of $RSS(R, C)$ is equal to the degrees of freedom for the error term, $N - J - K + 1$, times the expectation of the mean squared error (MSE) $\sigma_{2w,\varepsilon}^2$, since the $RSS(R, C)$ is always the SS for error as shown in Table 1.

Since there is no interaction term for the reader and case effects, the Type II and Type III SS are the same as those given in Table 1. The Type I SS depends on the order of terms added to the model. The SS of the last term added to the model and the error term will be the same as the Type II/III SS. We can compute the MS by dividing the SS with the degrees of freedom for each source of variation. The degrees of freedom for reader, case, and error are $J - 1$, $K - 1$ and $N - J - K + 1$, respectively. The expectation of MS can be derived from the expectation of RSS. Then, the variance components can be estimated by the method of moments. The estimates of variance components and the variance of WRBM differences can be expressed as linear combinations of the RSS. The coefficients in the linear combinations are shown in Table 4. For example, the variance estimate of WRBM differences by using Type I SS method with the reader effect adding to the model first is

$$\hat{V}_{WR}^{12} = \frac{1}{N_{1,R}} RSS(1) + \frac{N_{1,R} - N_{1,C}}{(N-J)N_{1,R}} RSS(R) + \frac{N_{1,C} - N + 1}{(N-J-K+1)N_{1,R}} RSS(R, C), \quad (4)$$

where $N_{1,C} = N - \frac{\sum_k (\sum_j n_{jk})^2}{N}$ and $N_{1,R} = N - \frac{\sum_j (\sum_k n_{jk})^2}{N}$ are the coefficients for the reader and case variance components in $E[RSS(1)]$, as given in the first row of Table 3. This equation shows how the coefficients of the RSS given in Row 4 of Table 4 combine to yield $\hat{V}_{WR}^{12}$.

### *2.2 Using three-way mixed effect ANOVA to estimate the variance of BRBM*

We cannot use a two-way ANOVA model to fit BRBM differences, as there are typically more than two readers in an MRMC study, and the pairs of readers cannot be assumed as independent levels for an effect. To estimate the variance of BRBM differences, we build a three-way mixed effect ANOVA model for the scores $X_{ijk}$ :

$$X_{ijk} = \mu + m_i + R_j + C_k + RC_{jk} + mR_{ij} + mC_{ik} + \varepsilon_{ijk}, \quad (5)$$

where $\mu$ denotes the grand mean, and $m_i$ represents the fixed effect for modality ($m_1 + m_2 = 0$). The other variables are the two random effects, reader and case, the interactions, and the random error. The random effects are independent and normally distributed with mean zero and variances given by $\sigma_R^2$, $\sigma_C^2$, $\sigma_{RC}^2$, $\sigma_{mR}^2$, $\sigma_{mC}^2$, $\sigma_\varepsilon^2$. The variance of a BRBM difference can be derived as:

$$V_{BR}^{12} = Var\left[D_{jj',k}^{12}\right] = 2\sigma_R^2 + 2\sigma_{RC}^2 + 2\sigma_{mR}^2 + 2\sigma_{mC}^2 + \sigma_\varepsilon^2 \quad (6)$$

The model structure is the same as what we used for the fully-crossed study. Like the SS for the two-way ANOVA model in Table 1, we define the three types of SS for the three-way ANOVA model with no replicates (Table 5). To compute Type I SS for all seven sources of variation, we need to fit seven multi-variable linear regressions. We need ten models for Type II SS and seven models for Type III SS. The SS for the error and the last term of the model are the same for all types of SS. Type II and Type III SS share the same SS for all the two-way interaction terms. To compute the Type III SS for the main effects, we need to fit the models that include the interaction terms containing the main effect without including the main effects themselves (shown in the first three rows of the last column in Table 5). This causes the estimation of the SS unreliable, which we will show and discuss in the results section. When computing Type I SS, there are only two possible orders of effects for the two-way ANOVA model. However, there are many more possible orders to add the effects for a three-way ANOVA model. As shown in the first two columns of Table 5, we choose to add the two random effects before including the modality effect and the interaction terms. One approach adds the reader effect first and the other adds case effect first. They share the same SS for the modality effect and all the two-way interaction terms. Each of the two columns sums up to $RSS(1)$.

As the number of terms in the multi-variable linear regression increases, it is impossible to derive the explicit forms of the RSS and their expectations. The full model can be express in the following matrix form:

$$\boldsymbol{X} = \boldsymbol{Z\beta} + \boldsymbol{\varepsilon} = \boldsymbol{Z_1}\beta_1 + \boldsymbol{Z_r\beta_r} + \boldsymbol{Z_c\beta_c} + \boldsymbol{Z_m\beta_m} + \boldsymbol{Z_{rc}\beta_{rc}} + \boldsymbol{Z_{rm}\beta_{rm}} + \boldsymbol{Z_{cm}\beta_{cm}} + \boldsymbol{\varepsilon} \quad (7)$$

where $\boldsymbol{\varepsilon} \sim \boldsymbol{N}(\boldsymbol{0}, \sigma_\varepsilon^2 \boldsymbol{I}_{N\times N})$ is the random error and $\boldsymbol{X}$ is a column vector for the observations (scores). The total number of observations is $N = \sum_i \sum_j \sum_k n_{ijk}$, where $n_{ijk} = 1$ when the reader $j$ read case $k$ under modality $i$. $\boldsymbol{Z}$ denotes the design matrix with $N$ rows and $N_\beta = 1 + J + K + I + JK + IJ + IK$ columns, one column for each level of an effect, organized by effect type: reader, case, modality, and interactions. $\boldsymbol{\beta}$ is a column vector with $N_\beta$ rows.

Equation (7) shows how the design matrix $\boldsymbol{Z}$ (and similarly, the vector $\boldsymbol{\beta}$) can be partitioned into smaller design matrices by its columns. $\boldsymbol{Z_1}$ is a column vector with all the elements as 1, indicating the design matrix for the intercept. The rest of the design matrices are for the six effects in the model with the number of columns equal to the number of levels for each effect. The elements in the design matrices are either zero or one and the row sum of each design matrix is one. For example, $\boldsymbol{Z_r}$ is $N \times J$, and $\{\boldsymbol{Z_r}\}_{nj} = 1$ means the $n$th observation in $\boldsymbol{X}$ is read by reader $j$ and the rest of elements in the $n$th row of $\boldsymbol{Z_r}$ should be zero. To compute the RSS of the full model, we need to estimate the effects $\boldsymbol{\beta}$ and calculate the residuals. When $\boldsymbol{Z^T Z}$ is invertible, the least squares solution to Equation (7) is $\widehat{\boldsymbol{\beta}} = (\boldsymbol{Z}^T\boldsymbol{Z})^{-1}\boldsymbol{Z}^T\boldsymbol{X}$, and we estimate $\boldsymbol{X}$ with $\widehat{\boldsymbol{X}} = \boldsymbol{Z}\widehat{\boldsymbol{\beta}} = \boldsymbol{Z}(\boldsymbol{Z}^T\boldsymbol{Z})^{-1}\boldsymbol{Z}^T\boldsymbol{X}.$ To simplify notation, let $\boldsymbol{H} = \boldsymbol{Z}(\boldsymbol{Z}^T\boldsymbol{Z})^{-1}\boldsymbol{Z^T}$denote the hat matrix. Given $\boldsymbol{H}$ is idempotent and $\boldsymbol{Z}$ is invariant under $\boldsymbol{H}$,

$$\begin{aligned} RSS(full) &= \left(\boldsymbol{X} - \widehat{\boldsymbol{X}}\right)^T \left(\boldsymbol{X} - \widehat{\boldsymbol{X}}\right) \\ &= \boldsymbol{X}^T(\boldsymbol{I} - \boldsymbol{H})\boldsymbol{X} \\ &= (\boldsymbol{Z\beta} + \boldsymbol{\varepsilon})^T(\boldsymbol{I} - \boldsymbol{H})(\boldsymbol{Z\beta} + \boldsymbol{\varepsilon}) \\ &= \boldsymbol{\varepsilon}^T(\boldsymbol{I} - \boldsymbol{H})\boldsymbol{\varepsilon}. \end{aligned} \tag{8}$$

Since $\boldsymbol{\varepsilon} \sim \boldsymbol{N}(\boldsymbol{0}, \sigma_\varepsilon^2 \boldsymbol{I}_{N\times N})$, $E[\boldsymbol{\varepsilon\varepsilon}^T] = \sigma_\varepsilon^2 \boldsymbol{I}_{N\times N}$. Then, the expected RSS for the full model is

$$\begin{aligned} E[RSS(full)] &= E[\boldsymbol{\varepsilon}^T(\boldsymbol{I} - \boldsymbol{H})\boldsymbol{\varepsilon}] \\ &= E[tr(\boldsymbol{\varepsilon}^T(\boldsymbol{I} - \boldsymbol{H})\boldsymbol{\varepsilon})] \\ &= E\left[tr\left((\boldsymbol{I} - \boldsymbol{H})\boldsymbol{\varepsilon\varepsilon}^T\right)\right] \\ &= tr\left((\boldsymbol{I} - \boldsymbol{H})E[\boldsymbol{\varepsilon\varepsilon}^T]\right) \end{aligned} \tag{9}$$

$$= tr(\boldsymbol{I} - \boldsymbol{H})\sigma_\varepsilon^2.$$

When the study is fully crossed, that is, $N = IJK$ , we can keep all $1 - J - K - I + JK + IJ + IK$ linearly independent columns in $\boldsymbol{Z}$ so that $\boldsymbol{Z^T Z}$ is full rank and $tr(\boldsymbol{I}_{N\times N} - \boldsymbol{H}) = IJK - IJ - IK - JK + I + J + K - 1$, which is the degrees of freedom for the error. When the data are from an arbitrary study design, there are no simple rules to extract linearly independent columns in $\boldsymbol{Z}$. To get the hat matrix, we do a QR decomposition for the design matrix, $\boldsymbol{Z} = \boldsymbol{QR}$, where $\boldsymbol{Q}$ is an orthogonal matrix and $\boldsymbol{R}$ is an upper triangular matrix. Let $\boldsymbol{Q_1}$ denote first $p$ linearly independent columns in $\boldsymbol{Q}$, and let $\boldsymbol{R_1}$ denote the first $p$ rows in $\boldsymbol{R}$. Then $\boldsymbol{Z} = [\boldsymbol{Q_1}, \boldsymbol{Q_2}]\begin{bmatrix}\boldsymbol{R_1}\\ \boldsymbol{R_2}\end{bmatrix} \approx \boldsymbol{Q_1 R_1}$ as $\boldsymbol{R_2} \approx \boldsymbol{0}$, and

$$\begin{aligned}\boldsymbol{H} &\approx \boldsymbol{Q_1 R_1}(\boldsymbol{R_1^T Q_1^T Q_1 R_1})^{-1}\boldsymbol{R_1^T Q_1^T} \\ &= \boldsymbol{Q_1 R_1 R_1^{-1}}(\boldsymbol{R_1^T})^{-1}\boldsymbol{R_1^T Q_1^T} \\ &= \boldsymbol{Q_1 Q_1^T}\end{aligned} \tag{10}$$

Therefore, $RSS(full) = \boldsymbol{X}^T(\boldsymbol{I}_{N\times N} - \boldsymbol{Q_1 Q_1^T})\boldsymbol{X}$ and $E[RSS(full)] = tr(\boldsymbol{I}_{N\times N} - \boldsymbol{Q_1 Q_1^T})\sigma_\varepsilon^2 = (N - p)\sigma_\varepsilon^2$.

The RSS and its expectation for the other models can be derived similarly. Supplementary Material A take model R, C, M, RC as an example. Based on those, we compute the different types of SS, their expectation, and consequently, all the variance components in the three-way mixed effect ANOVA model.

### *2.3 Simulation study*

The following normal-inverse-gamma (Normal-IG) hierarchical simulation model(Wen and Gallas 2022) can be used to simulate MRMC agreement data:

$$X_{ijkl} = \mu + \tau_i + [RC]_{jk} + [\tau RCE]_{ijkl},$$

$$[RC]_{jk} | \, R_j, \, C_k \sim N(C_k, R_j), \tag{11}$$

$$[\tau RCE]_{ijkl} | \, [\tau R]_{ij}, [\tau C]_{ik} \sim N([\tau C]_{ik}, [\tau R]_{ij}).$$

Also, the case effect $C_k \sim N(0, \sigma_C^2)$, the reader effect $R_j \sim \text{InvG}(\alpha_R, \beta_R)$, the modality-case interaction $[\tau C]_{ik} \sim N(0, \sigma_{\tau C}^2)$, and the modality-reader interaction $[\tau R]_{ij} \sim \text{InvG}(\alpha_{\tau R}, \beta_{\tau R})$. The reader-case interaction is conditionally normal with the case effect equal to the mean and the reader effect equal to the variance. Similarly, the four-way interaction is conditionally normal with the modality-case interaction equal to the mean and the modality-reader interaction equal to the variance. By using this hierarchical simulation model, we can simulate readers with different reader variability where differences may be expected due to differences in reader's experience level, expertise, and training. The MRMC data simulated by this non-linear model can also help us to test whether the ANOVA estimation can be robust beyond normally distributed data, as assumed in the ANOVA model.

In the simulation studies here, we focus on validating the variance estimates of WRBM and BRBM differences for arbitrary study designs. In the real data example, we will demonstrate a within-modality comparison.

Given the simulation model, the true value for the mean and variance of the WRBM and BRBM differences can be derived as following:

$$\overline{D}_{WR}^{12} = \tau_i - \tau_{i'}, \tag{12}$$

$$V_{WR}^{12} = 2\sigma_{\tau C}^2 + \frac{2\beta_{\tau R}}{\alpha_{\tau R} - 1}, \tag{13}$$

$$\overline{D}_{BR}^{12} = \tau_i - \tau_{i'}, \tag{14}$$

$$V_{BR}^{12} = 2\sigma_{\tau C}^2 + \frac{2\beta_{\tau R}}{\alpha_{\tau R} - 1} + \frac{2\beta_R}{\alpha_R - 1}. \tag{15}$$

We simulate $T = 1000$ MRMC studies for each parameter setting and denote the variance estimates of WRBM and BRBM differences for each study $t$ by $\hat{V}_{WR,t}^{12}$and $\hat{V}_{BR,t}^{12}$. The estimation performance is assessed by relative bias and the coefficient of variation (CV):

$$relative_bias\left(\hat{V}_{WR}^{12}\right) = \frac{\frac{1}{T}\sum_t \hat{V}_{WR,t}^{12} - V_{WR}^{12}}{V_{WR}^{12}}, \tag{16}$$

$$CV\left(\hat{V}_{WR}^{12}\right) = \frac{SD\left(\hat{V}_{WR,t}^{12}\right)}{V_{WR}^{12}}, \tag{17}$$

where $SD()$ denotes the sample standard deviation across the $T$ studies.

The following four sets of parameters were tested:

(1) Different number of readers reading a fixed number of cases: $J = 4,5, \dots, 10, 15, \dots 50, K = 30,50,100$. The reader and case variability parameters were $\alpha_R = 6$, $\sigma_C^2 = 0.4$. The study design was the design with random workload by batch of cases, as shown in Figure 1.b. Every 10 cases were grouped into a batch. We randomly removed 40% of the reader-batch combinations from a fully-crossed study. Thus, it had 40% missing data.

(2) Different number of cases read by a fixed number of readers: $J = 5,10,15, K = 30,40, \dots, 150$. The reader and case variability parameters were $\alpha_R = 6$, $\sigma_C^2 = 0.4$. The study design had 40% missing data in a design with random workload by batch of cases.

(3) Different study designs with different amounts of missing data. The number of readers and cases were $J = 5, K = 50$, and the reader and case variability parameters were $\alpha_R = 6$, $\sigma_C^2 = 0.4$. We explored 0, 10%, 20%, 30%, 40% missing data in the four types of arbitrary study design as shown in Figure 1.a-Figure 1.d.

(4) Different levels of reader and case variabilities. The reader and case variability parameters explored were $(\alpha_R, \sigma_C^2) \in \{3,4,6,11,21\} \times \{0.1,0.2,0.4, 2/3, 1\}$. The number of readers and cases were $J = 5$, $K = 50$. The study design had 40% missing data in a design with random workload by batch of cases.

In all the parameter sets, $\beta_R = \beta_{\tau R} = 1$, so that the reader variability is only affected by $\alpha_R (= \alpha_{\tau R})$. When $\alpha_R = \alpha_{\tau R} = 6$ and $\sigma_C^2 = 0.4$, the contributions to $V_{BR}^{12}$ from reader and case variability are both equal to 0.8 ($= 2\sigma_{\tau C}^2 = \frac{2\beta_{\tau R}}{\alpha_{\tau R}-1} + \frac{2\beta_R}{\alpha_R - 1}$, Equation (15)).

### *2.4 Real data example*

In this real data example, we estimate both the BRBM and BRWM limits of agreement by using data from the High-Throughput Truthing (HTT) project.(Dudgeon et al. 2021) The pathologists who participated in the project estimated the density of stromal tumor-infiltrating lymphocytes (sTILs) of each region of interest (ROI) pre-selected from 64 tissue slides from breast cancer patients. Ten ROIs were selected from each tissue slide, but not all the ROIs are deemed evaluable for sTIL evaluation. The pathologists review the ROIs through two digital platforms: caMicroscope(Saltz et al. 2017) and PathPresenter.(*PathPresenter* n.d.) Not all the pathologists finished reviewing all the ROIs on both the platforms. In total, there are 5 604 sTILs scores from 25 pathologists for 594 ROIs. As a density, each score is within the range of 0-100. In this example, we assume the ROIs are independent from each other and ignore the correlations among the ROIs that are selected from the same slides.

To compare the performance of the pathologists across different platforms, we estimated the BRBM limits of agreement by using the three-way ANOVA method illustrated above. The BRBM LOA is compared with the BRWM LOA for each individual platform. Supplementary Material C shows how to use both ANOVA models to estimate the variance of BRWM.

## 3. Results

### *3.1 Simulation Studies*

Figure 2, Figure 3 and Figure 4 demonstrate the relative bias and CV for the variance estimates of BRBM differences for the first three sets of parameters. The results for the WRBM differences and the fourth set of parameters can be found in the Supplementary Material B. In all three figures, plots in the top row describe the relative bias observed and the plots in the bottom row present the CV. The different shapes of points illustrate the different types of SS used for the ANOVA method as given in the legend. For any parameter setting, the relative bias and CV observed using the Type III SS method is larger than that estimated by the Type I or Type II SS methods, while the results from the Type I and Type II SS methods are relatively close to each other.

The large relative bias and CV for variance estimates by Type III SS methods is mainly caused by bias in estimating the variance components for the reader effect. As shown in Table 5, Type II and Type III SS share the same SS for the interaction terms and differ only for the main effects. The linear regressions for estimating Type III SS for the main effects include interaction terms involving the main effects. Thus, the marginal variation in the data that is left to be explain by the main effect is small. This causes the Type III SS for the main effects to be smaller than the Type II SS. However, the estimates for $\sigma_R^2$ and $\sigma_C^2$ using Type III methods are often larger than those using Type II methods. This is because fewer variance components contribute to the expectation of Type III SS. The expectation of RSS is a linear combination of the variance components for the effects that are not added to the linear regression. Thus, the expectation of Type III SS for the reader effect is only related to $\sigma_R^2$ and $\sigma_\varepsilon^2$; whereas the expectation of Type II SS for the reader effect is a linear combination of $\sigma_R^2$, $\sigma_{RC}^2$, $\sigma_{mR}^2$ and $\sigma_\varepsilon^2$.This results in the large bias and uncertainty estimating the variance of BRBM differences using Type III SS method.

For the Type I and Type II SS methods, as we increase the size of the MRMC study, the CVs of the variance estimates of BRBM differences in both Figure 2 and Figure 3 decrease smoothly. It might not be clear in the figures under the low resolution, but when the sample size of the study is small, say five readers and 30 cases, the CVs of variance estimates of BRBM differences by using Type II SS method are 11% ($= \frac{0.23-0.207}{0.207} \times 100\%$) larger than those using Type I SS method (Figure 3). Similarly, when the percentage of missing data is at 40% in the design with random workload, the CV for Type II SS method is 9% ($= \frac{0.175-0.16}{0.16} \times 100\%$) larger than that for Type I SS method (Figure 4). This illustrates that the variance estimate using Type I SS method has less uncertainty compared to the variance estimates from the Type II and Type III methods, especially for the small MRMC study.

When comparing different study designs, we observe the uncertainty in variance estimates of BRBM differences to be more sensitive to the sample size for the designs that generate more unbalanced data, like the design with random workload (first column in Figure 4 arising from the study design in Figure 1.a) and the design that groups cases into batch (second column in Figure 4 arising from the study design in Figure 1.b). When we set the percentage of missing data beyond 40%, the variance of BRBM differences is not always estimable. This happens when the number of observations $N$ is smaller than $p$, the number of independent columns in $\boldsymbol{Z}$. In these cases, $E[RSS(full)] = 0$, and the variance component for the error term is not estimable. To avoid this situation, the degree of freedom for the error, which is the number of observations minus the sum of degrees of freedoms for other sources of variation, needs to be greater than 0, that is,

$$N - IJ - IK - JK + I + J + K - 1 > 0. \quad (18)$$

When we use $w$ to denote the percentage of missing data and focus on comparing two modalities, $I = 2$, then $N = 2(1 - w)JK$. Thus, the percentage of missing data that our method can treat is limited by

$$w < \frac{(J-1)(K-1)}{2JK}. \quad (19)$$

This is a conservative limit as the degrees of freedoms for the sources of variation are calculated under the fully-crossed study assumption. If the missing is structured, for example when the missing is paired across the two modalities, the degrees of freedom for the reader and case interaction will be smaller than $(J - 1)(K - 1)$. We can allow smaller number of observations and still get positive degrees of freedom for the error. In other words, given the same percentage of missing data, the degrees of freedom for the error will be larger and more stable for the study design with paired missing data. This can also explain why the CVs in the last two columns of plots in Figure 4 do not change much as we increase the percentage of missing data.

### *3.2 Real data example*

Table 6 lists the MRMC variance estimates and LOA for BRBM and BRWM differences for the two evaluation platforms, caMicroscope and PathPresenter, using different types of SS. The first row of the top half of the table presents the variance estimates of BRBM differences, $\hat{V}_{BR}^{12}$. The mean difference between pairs of readers evaluating the same ROI from different platforms is 1.378. Thus, the estimated BRBM LOA is $1.378 \pm 1.96 * \sqrt{\hat{V}_{BR}^{12}}$, as shown in the first row of bottom half of Table 6.

In current practice, the consensus measurements for each case are generated by averaging over the measurements from multiple readers, and different modalities are compared using the independent consensus measurements. If we generate consensus measurements for each case in each platform, the differences of the consensus measurements between the two platforms have mean 0.5413 and variance 56.19. Thus, the estimated limits of agreement between the two platform would be (-14.15, 15.23), which is much narrower compared to the BRBM LOA. This is because the reader variability is not considered when estimating the variance of the difference between consensus measurements.

The second and third rows of the top half of Table 6 show the variance estimates of BRWM differences for each platform, $\hat{V}_{BR}^{ii}$. As the order of the readers in the BRWM differences does not matter, the mean BRWM difference is 0 by default, and the estimated BRWM limits of agreement will be $\pm 1.96 * \sqrt{\hat{V}_{BR}^{ii}}$. The fact that the BRWM LOA for caMicroscope covers the BRBM LOA indicates the agreement between readers across different platforms is comparable to the agreement within the platform. Therefore, the data do not demonstrate significant impact of using different platforms to evaluate the sTILs density.

When comparing the estimates by different methods, we see that the variance estimate of BRBM differences by Type III SS method is very different from those by Type I and II SS methods. This is consistent from what we observed in the simulation study. When considering the variance estimates of BRWM differences, it is only determined by $\hat{\sigma}_{2w,R}^2$ and $\hat{\sigma}_{2w,\varepsilon}^2$, as shown in Equation (S5) in Supplementary Material C. From Table 4 we know that those two estimates are the same for Type I SS method with case first and Type II/III SS method. Therefore, the variance estimates for BRWM differences are the same for those three methods.

## 4. Discussions and Conclusion

In this work, we focus on generalizing MRMC estimates of the LOA to treat arbitrary MRMC study designs. For the between modality LOA, a three-way mixed effect ANOVA model is applied to estimate the variance of BRBM differences, while a two-way random effect ANOVA model is used to estimate the LOA arising from WRBM differences. For the within-modality comparison, we explained how to use the two-way or three-way to estimate the variance of BRWM differences and that they give the same result. It's worth pointing out that to estimate WRWM LOA, some of the readers need to read some of the cases twice. Then we can compute the WRWM differences $D_{jj,k}^{ii} = X_{ijkl} - X_{ijkl'}$ and use a two-way random effect ANOVA model to decompose the total variation in the data. The variance estimate for WRWM differences will be $\hat{V}_{BR}^{ii} = \hat{\sigma}_{2w,\varepsilon}^{2}$. Therefore, as summarized in Figure 5, three of the four types of MRMC LOA (WRWM, WRBM, and BRWM) can be estimated by using a two-way random effect ANOVA model. For the within-reader comparison, the model is built on the computed difference score. To estimate the variance of the BRWM difference, the model takes the raw score as the response variable. The three-way random effect ANOVA model for estimating the variance of BRBM differences is also built on the raw score. This is because difference pairs of readers cannot be considered as independent levels of an effect.

When comparing the CV of the variance estimate for BRBM differences for different size of the MRMC study, we notice that the CVs for Type I/II methods in Figure 2 seem to plateau as the number of readers reaches 20, while the CVs in Figure 3 keep decreasing as the number of cases increase. This may be due to the nature of the simulation model. To better understand this issue, one of our future works is to estimate the precision of the variance estimates. Then we can further explore the relationship between the precision of the variance estimate and the size of the simulation study.

There are some other limitations in our simulation studies. We do not observe a difference in the performance characteristics of the variance estimation for the Type I method when the reader was added first compared to when the case effect was added first. This is not to say that the order of the terms does not matter. We only compare two orders of effects for the three-way ANOVA model. As mentioned in the method section, there are many other orders for the seven effects. More simulations are needed to explore which order of effect has the best performance of estimation. In addition, we observe that the estimation for the variance component $\hat{\sigma}_R^2$ and $\hat{\sigma}_{mR}^2$ could be negative in some of the Monte Carlo trials. This happens more often when the number of readers is small, or the reader variability is small. Further studies need to be done to better understand why this happen and what is the best strategy to correct the negative estimation.

To run the simulation study and real data example, we developed software to estimate the variance of WRBM and BRBM differences. Some existing functions in R packages can provide ANOVA tables for unbalanced data. For example, the *"aov"* function in the *"stats"* package(R Core Team 2021) returns the ANOVA table for the Type I SS and the *"Anova"* function in the *"car"* package(Fox and Weisberg 2018) returns the ANOVA table for the Type II or III SS. However, these existing functions do not provide estimates of the variance components for the random effects. We developed software in R to output the different types of SS, the estimates for the variance components, and the variance of different types of MRMC differences. The Type I, II and III SS generated from our functions are the same as those output from the existing functions mentioned. Our software is available at https://github.com/DIDSR/ANOVA.MRMC.LOA.

Among the three types of ANOVA methods discussed and compared here, the Type I ANOVA is the most efficient method, considering the computational cost. Table 7 shows the processing time for Type I, II and III SS methods from MRMC studies of different sizes. The QR decomposition is the most time-consuming part of the code. For the Type I SS method, the terms are added to the model in sequence. Therefore, only one QR decomposition is required for the full model and the Q matrices for the six smaller regression models can be extracted from the columns of the Q matrix for the full model. In contrast, for the Type II SS method we need seven QR decompositions to compute all the RSS, and six QR decompositions are needed for the Type III method. Thus, the processing time for Type II or III SS methods are longer than that for Type I SS method. When dealing with larger MRMC studies, the difference in processing time is even larger.

In summary, we validated the variance estimates of WRBM and BRBM differences for arbitrary study designs at different parameter settings. We observed that the variance estimates using Type III SS method were positively biased with large uncertainty in our simulation results. The variance estimates from both the Type I and II SS were observed to be unbiased as expected. When the size of the MRMC study is small or the number of missing values is large, the uncertainty for the Type I SS method is about 10% smaller than that for the Type II SS method. Also, the Type I SS method is more time efficient than the other two methods. Therefore, we believe Type I SS method should be considered first when estimating the limits of agreement for an MRMC study with arbitrary study design.

**Declaration of Interest Statement**

The authors report there are no competing interests to declare

Table 1 Different types of SS for two-way ANOVA model with no replicates

| SS | Type I – reader first | Type I – case first | Type II | Type III |
|---|---|---|---|---|
| **Reader** | $RSS(1) - RSS(R)$ | $RSS(C) - RSS(R,C)$ | $RSS(C) - RSS(R,C)$ | $RSS(C) - RSS(R,C)$ |
| **Case** | $RSS(R) - RSS(R,C)$ | $RSS(1) - RSS(C)$ | $RSS(R) - RSS(R,C)$ | $RSS(R) - RSS(R,C)$ |
| **Error** | $RSS(R,C)$ | $RSS(R,C)$ | $RSS(R,C)$ | $RSS(R,C)$ |
| **Total** | $RSS(1)$ | $RSS(1)$ | $RSS(R) + RSS(C) - RSS(R,C)$ | $RSS(R) + RSS(C) - RSS(R,C)$ |

Table 2 Different types of comparisons in an MRMC study

| Comparison | Within-Reader Within-Modality (WRWM) | Between-Reader Within-Modality (BRWM) | Within-Reader Between-Modality (WRBM) | Between-Reader Between-Modality (BRBM) |
|---|---|---|---|---|
| **Difference** | $D_{jj,k}^{ii} = X_{ijkl} - X_{ijkl'}$* | $D_{jj',k}^{ii} = X_{ijk} - X_{ij'k}$ | $D_{jj,k}^{12} = X_{1jk} - X_{2jk}$ | $D_{jj',k}^{12} = X_{1jk} - X_{2j'k}$ |
| **LOA** | $\bar{D}_{WR}^{ii} \pm 1.96\sqrt{V_{WR}^{ii}}$ | $\bar{D}_{BR}^{ii} \pm 1.96\sqrt{V_{BR}^{ii}}$ | $\bar{D}_{WR}^{12} \pm 1.96\sqrt{V_{WR}^{12}}$ | $\bar{D}_{BR}^{12} \pm 1.96\sqrt{V_{BR}^{12}}$ |

*$l$ and $l'$ in the subscript denotes the different replicates when a reader reads the same case within a single modality.

Table 3 Residual sum of squares and their expectation for different linear models.

| | **Model Structure** | **RSS *** | **E(RSS) **** |
|---|---|---|---|
| **Model for RSS(1)** | $D_{jj,k}^{12} = \mu + \varepsilon_{jk}$, $\varepsilon_{jk} \sim N(0, \sigma_{2w,\varepsilon}^2)$ | $\sum_j \sum_k (d_{jk} - d_{..})^2 n_{jk}$ | $N_{1,C}\sigma_{2w,C}^2 + N_{1,R}\sigma_{2w,R}^2 + (N-1)\sigma_{2w,\varepsilon}^2$ |
| **Model for RSS(R)** | $D_{jj,k}^{12} = \mu + r_j + \varepsilon_{jk}$, $\varepsilon_{jk} \sim N(0, \sigma_{2w,\varepsilon}^2)$, $\sum_j r_j = 0$ | $\sum_j \sum_k (d_{jk} - d_{j.})^2 n_{jk}$ | $(N-J)(\sigma_{2w,C}^2 + \sigma_{2w,\varepsilon}^2)$ |
| **Model for RSS(C)** | $D_{jj,k}^{12} = \mu + c_k + \varepsilon_{jk}$, $\varepsilon_{jk} \sim N(0, \sigma_{2w,\varepsilon}^2)$, $\sum_j c_j = 0$ | $\sum_j \sum_k (d_{jk} - d_{j.})^2 n_{jk}$ | $(N-K)(\sigma_{2w,R}^2 + \sigma_{2w,\varepsilon}^2)$ |
| **Model for RSS(R, C)** | $D_{jj,k}^{12} = \mu + r_j + c_k + \varepsilon_{jk}$, $\varepsilon_{jk} \sim N(0, \sigma_{2w,\varepsilon}^2)$, $\sum_j r_j = 0$, $\sum_k c_k = 0$ | $RSS(R,C)$*** | $(N-J-K+1)\sigma_{2w,\varepsilon}^2$ |

* $d_{..} = \frac{\sum_j \sum_k d_{jk} n_{jk}}{N}$, $d_{j.} = \frac{\sum_k d_{jk} n_{jk}}{\sum_k n_{jk}}$, $d_{.k} = \frac{\sum_j d_{jk} n_{jk}}{\sum_j n_{jk}}$

** $N = \sum_j \sum_k n_{jk}$, $N_{1,C} = N - \frac{\sum_k (\sum_j n_{jk})^2}{N}$, $N_{1,R} = N - \frac{\sum_j (\sum_k n_{jk})^2}{N}$

*** Solved indirectly. Refer to the text around Equation 3.

Table 4 Estimates of variance components and the variance of WRBM differences in form of linear combinations of the RSS

| SS | Variance Estimates | Coefficients for the RSS | | | |
|---|---|---|---|---|---|
| | | $RSS(1)$ | $RSS(R)$ | $RSS(C)$ | $RSS(R,C)$ |
| **Type I - Reader First** | $\hat{\sigma}^2_{2w,R}$ | $\frac{1}{N_{1,R}}$ | $-\frac{N_{1,C}}{(N-J)N_{1,R}}$ | 0 | $\frac{N_{1,C}-N+1}{(N-J-K+1)N_{1,R}}$ |
| | $\hat{\sigma}^2_{2w,C}$ | 0 | $\frac{1}{N-J}$ | 0 | $-\frac{1}{N-J-K+1}$ |
| | $\hat{\sigma}^2_{2w,\varepsilon}$ | 0 | 0 | 0 | $\frac{1}{N-J-K+1}$ |
| | $\hat{V}^{12}_{WR}$ | $\frac{1}{N_{1,R}}$ | $\frac{N_{1,R}-N_{1,C}}{(N-J)N_{1,R}}$ | 0 | $\frac{N_{1,C}-N+1}{(N-J-K+1)N_{1,R}}$ |
| **Type I - Case First** | $\hat{\sigma}^2_{2w,R}$ | 0 | 0 | $\frac{1}{N-K}$ | $-\frac{1}{N-J-K+1}$ |
| | $\hat{\sigma}^2_{2w,C}$ | $\frac{1}{N_{1,C}}$ | 0 | $-\frac{N_{1,R}}{(N-K)N_{1,C}}$ | $\frac{N_{1,R}-N+1}{(N-J-K+1)N_{1,C}}$ |
| | $\hat{\sigma}^2_{2w,\varepsilon}$ | 0 | 0 | 0 | $\frac{1}{N-J-K+1}$ |
| | $\hat{V}^{12}_{WR}$ | $\frac{1}{N_{1,C}}$ | 0 | $\frac{N_{1,R}-N_{1,C}}{(N-K)N_{1,R}}$ | $\frac{N_{1,R}-N+1}{(N-J-K+1)N_{1,C}}$ |
| **Type II/III** | $\hat{\sigma}^2_{2w,R}$ | 0 | 0 | $\frac{1}{N-K}$ | $-\frac{1}{N-J-K+1}$ |
| | $\hat{\sigma}^2_{2w,C}$ | 0 | $\frac{1}{N-J}$ | 0 | $-\frac{1}{N-J-K+1}$ |
| | $\hat{\sigma}^2_{2w,\varepsilon}$ | 0 | 0 | 0 | $\frac{1}{N-J-K+1}$ |
| | $\hat{V}^{12}_{WR}$ | 0 | $\frac{1}{N-J}$ | $\frac{1}{N-K}$ | $-\frac{1}{N-J-K+1}$ |

Table 5 Different types of SS for three-way ANOVA model with no replicates

| SS | Type I – reader first | Type I – case first | Type II | Type III |
|---|---|---|---|---|
| **Reader** | $RSS(1) -$ $RSS(R)$ | $RSS(C) -$ $RSS(R, C)$ | $RSS(C, M, CM) -$ $RSS(R, C, M, CM)$ | $RSS(C, M, RC, RM,$ $RSS(full)$ |
| **Case** | $RSS(R) -$ $RSS(R, C)$ | $RSS(1) - RSS(C)$ | $RSS(R. M, RM) -$ $RSS(R, C, M, RM)$ | $RSS(R, M, RC, RM,$ $RSS(full)$ |
| **Modality** | $RSS(R, C) -$ $RSS(R, C, M)$ | $RSS(R, C) -$ $RSS(R, C, M)$ | $RSS(R. C, RC) -$ $RSS(R, C, M, RC)$ | $RSS(R, C, RC, RM,$ $RSS(full)$ |
| **Reader x Case** | $RSS(R, C, M) -$ $RSS(R, C, M, RC)$ | $RSS(R, C, M) -$ $RSS(R, C, M, RC)$ | $RSS(R, C, M, RM, C$ $RSS(full)$ | $RSS(R, C, M, RM, C$ $RSS(full)$ |
| **Reader x Modality** | $RSS(R, C, M, RC) -$ $RSS(R, C, M, RC, R$ | $RSS(R, C, M, RC) -$ $RSS(R, C, M, RC, R$ | $RSS(R, C, M, RC, C$ $RSS(full)$ | $RSS(R, C, M, RC, C$ $RSS(full)$ |
| **Case x Modality** | $RSS(R, C, M, RC, R$ $RSS(full)$ | $RSS(R, C, M, RC, R$ $RSS(full)$ | $RSS(R, C, M, RC, R$ $RSS(full)$ | $RSS(R, C, M, RC, R$ $RSS(full)$ |
| **Error** | $RSS(full)$ | $RSS(full)$ | $RSS(full)$ | $RSS(full)$ |

Table 6 Variance estimates and limits of agreement for BRBM and BRWM differences for HTT project dataset by using different types of SS.

| **Variance of Difference** | | **Type I – reader first** | **Type I – case first** | **Type II** | **Type III** |
|---|---|---|---|---|---|
| **BRBM** | caMicroscope - PathPresenter | 180.0 | 176.9 | 173.0 | 101.2 |
| **BRWM** | caMicroscope | 201.9 | 199.9 | 199.9 | 199.9 |
| | PathPresenter | 132.4 | 121.9 | 121.9 | 121.9 |

| **Limits of Agreement** | | **Type I – reader first** | **Type I – case first** | **Type II** | **Type III** |
|---|---|---|---|---|---|
| **BRBM** | caMicroscope - PathPresenter | (-24.92, 27.67) | (-24.69, 27.45) | (-24.40, 27.16) | (-18.33, 21.09) |
| **BRWM** | caMicroscope | (-27.85, 27.85) | (-27.71, 27.71) | (-27.71, 27.71) | (-27.71, 27.71) |
| | PathPresenter | (-22.55, 22.55) | (-21.64, 21.64) | (-21.64, 21.64) | (-21.64, 21.64) |

Table 7 Processing time for Type I, II, and III SS methods (one dataset).

| Processing Time (seconds) | Type I | Type II | Type III |
|---|---|---|---|
| $\boldsymbol{J = 10, K = 100}$ | 4.12 | 8.75 | 7.62 |
| $\boldsymbol{J = 10, K = 200}$ | 24.74 | 74.61 | 66.33 |

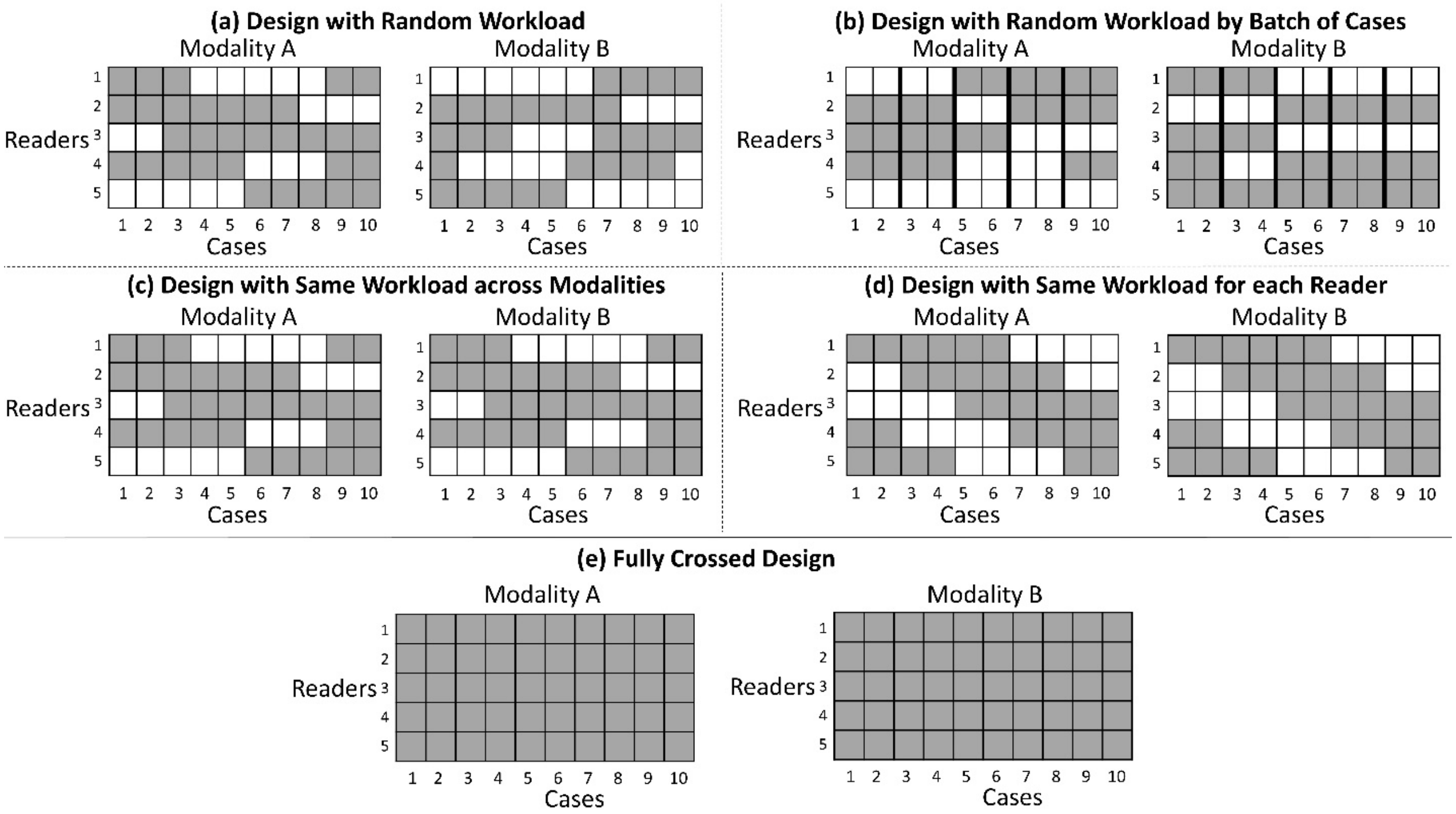

Figure 1 Different types of MRMC study designs with two reading modalities. Each row in each plot represents one reader; each column represents one case. An element is filled gray if the reader in that row reads the case in that column for that modality; otherwise, the element is white. When all the readers read all the cases on both modalities, the study is **(e)** fully crossed. Some examples of not fully-crossed, or arbitrary designs, are **(a)** design with random workload, **(b)** design that groups cases into batches and readers are assigned random batches, **(c)** design with readings paired across the modalities, and **(d)** design with each reader having the same workload, each case being read by the same number of readers and paired readings across the modalities.

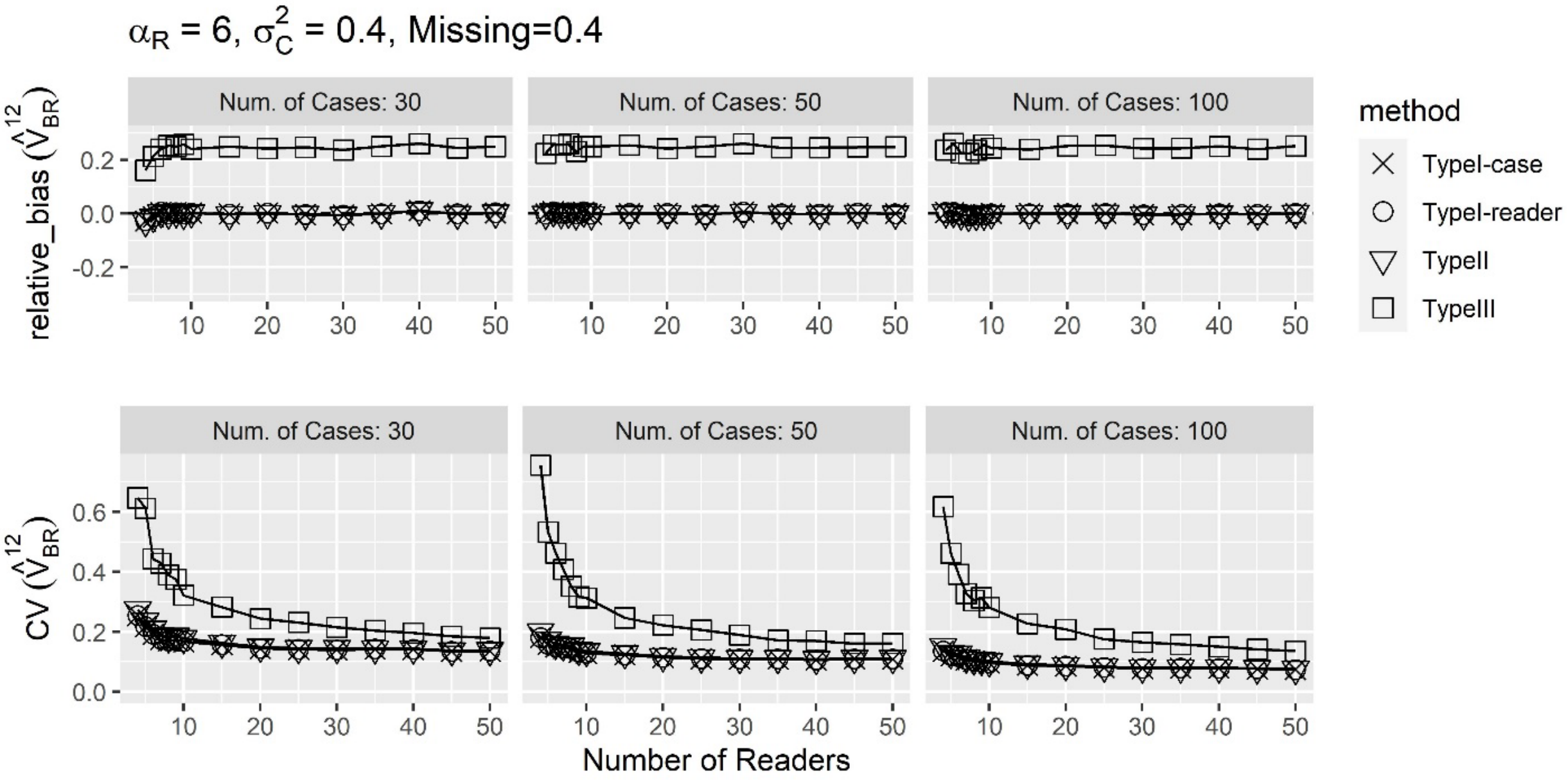

Figure 2 Relative bias and CV of variance estimates for BRBM differences as a function of the number of readers for different numbers of cases

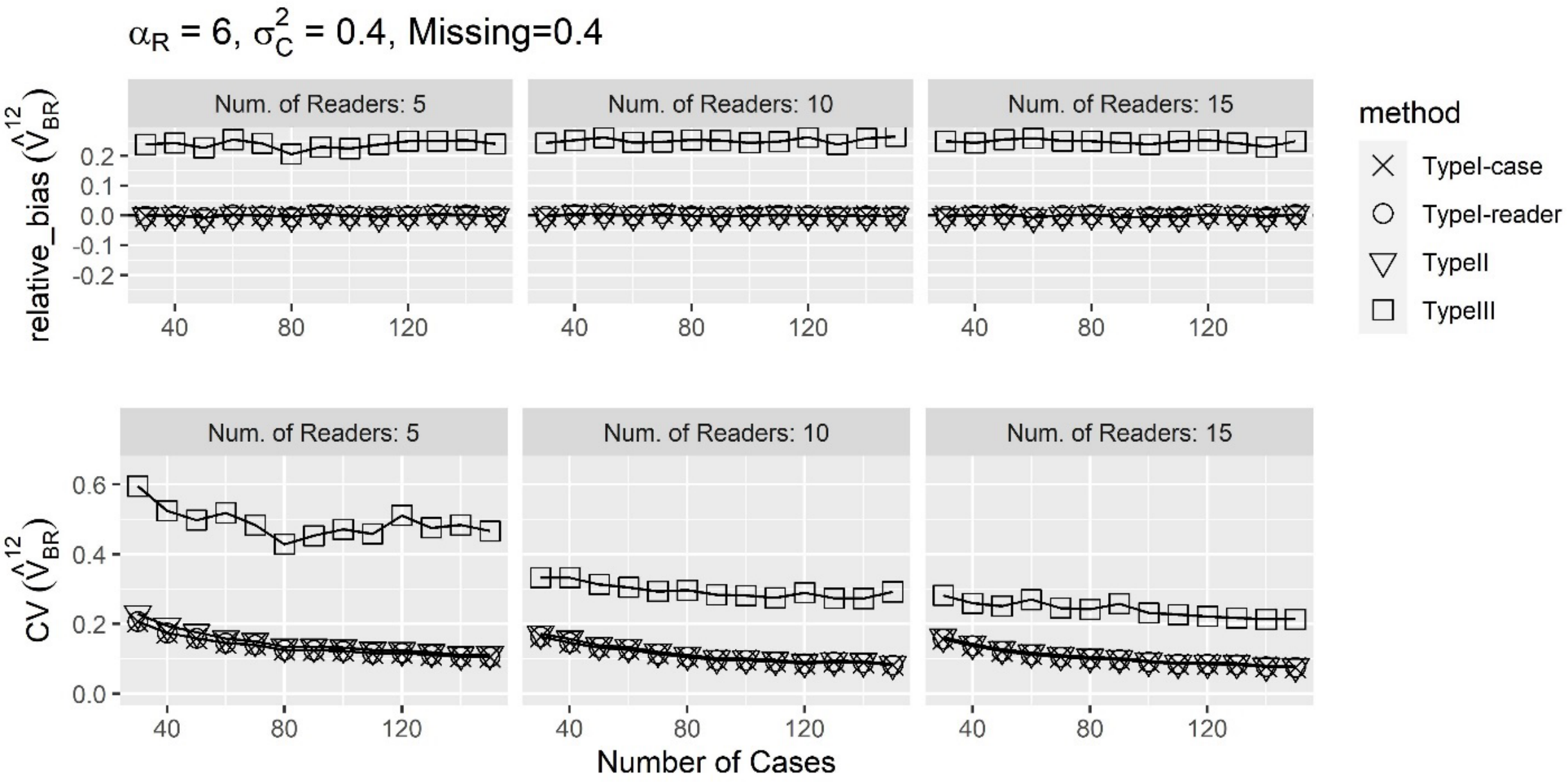

Figure 3 Relative bias and CV of variance estimates for BRBM differences as a function of the number of cases for different numbers of readers

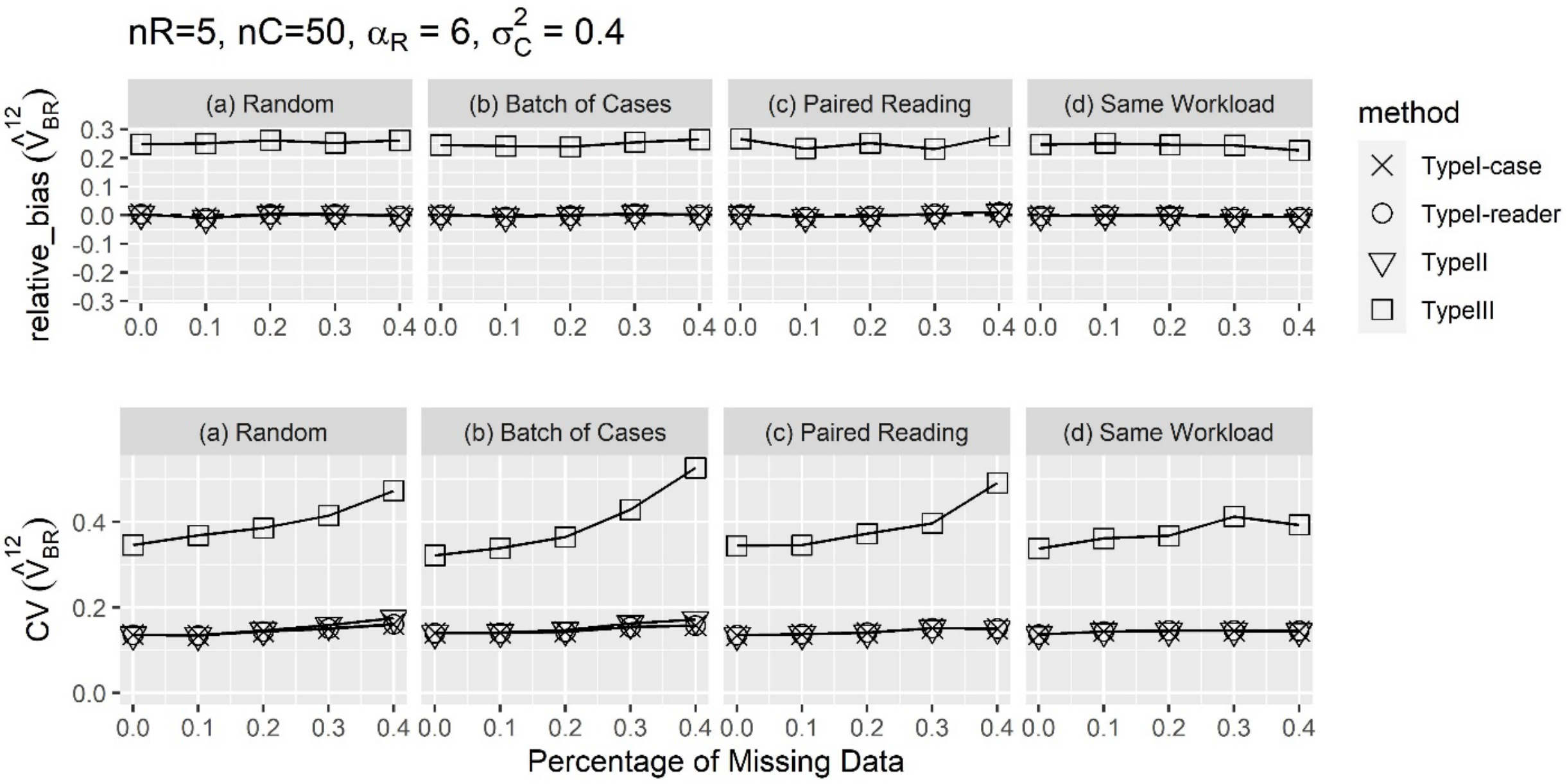

Figure 4 Relative bias and CV of variance estimates for BRBM differences as a function of the percent of missing values for different study designs

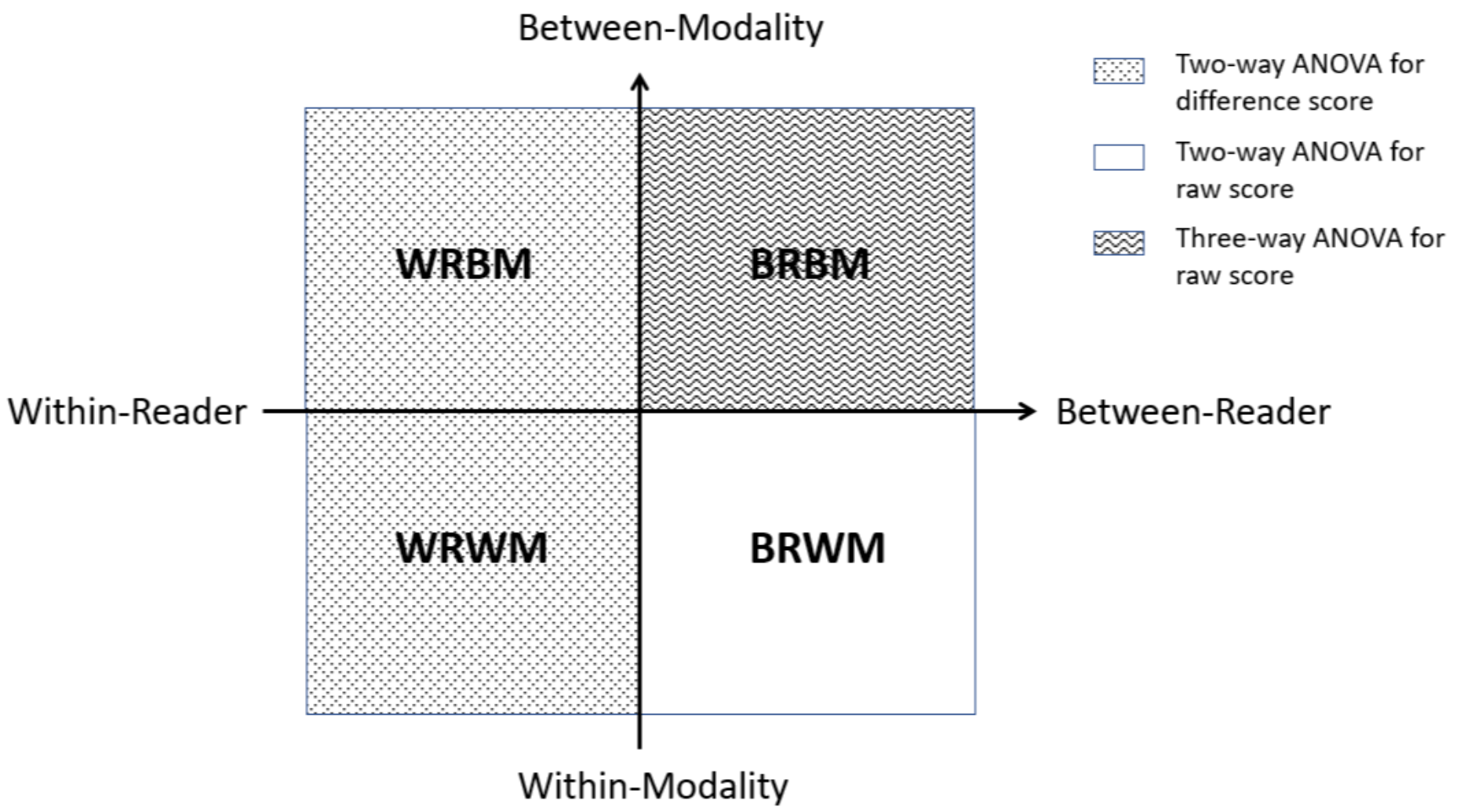

Figure 5 Different ANOVA models for estimating the variance of different types of MRMC difference score

## Figure Captions

Figure 1 Different types of MRMC study designs with two reading modalities. Each row in each plot represents one reader; each column represents one case. An element is filled gray if the reader in that row reads the case in that column for that modality; otherwise, the element is white. When all the readers read all the cases on both modalities, the study is **(e)** fully crossed. Some examples of not fully-crossed, or arbitrary designs, are **(a)** design with random workload, **(b)** design that groups cases into batches and readers are assigned random batches, **(c)** design with readings paired across the modalities, and **(d)** design with each reader having the same workload, each case being read by the same number of readers and paired readings across the modalities.

Figure 2 Relative bias and CV of variance estimates for BRBM differences as a function of the number of readers for different numbers of cases

Figure 3 Relative bias and CV of variance estimates for BRBM differences as a function of the number of cases for different numbers of readers

Figure 4 Relative bias and CV of variance estimates for BRBM differences as a function of the percent of missing values for different study designs

Figure 5 Different ANOVA models for estimating the variance of different types of MRMC difference score

**Supplementary Material**

## A. Residual Sum of Squares and their Expectation for the Smaller Models

Here we take model R, C, M, RC as an example. Its RSS, $RSS(R, C, M, RC)$, is used to compute Type I SS for reader-case interaction (shown at 4th row and the first two columns in Table 5) and Type II SS for modality (shown at 3rd row and 3rd column in Table 5). The design matrix for this model can be expressed as $\boldsymbol{Z}_{R*C+M} = [\boldsymbol{Z_1}, \boldsymbol{Z_r}, \boldsymbol{Z_c}, \boldsymbol{Z_m}, \boldsymbol{Z_{rc}}]$. The multi-variable linear regression can be expressed as

$$\begin{aligned} \boldsymbol{X} &= \boldsymbol{Z}_{R*C+M}\boldsymbol{\beta} + \boldsymbol{\varepsilon}_{R*C+M} \\ &= \boldsymbol{Z_1}\beta_1 + \boldsymbol{Z_r}\boldsymbol{\beta_r} + \boldsymbol{Z_c}\boldsymbol{\beta_c} + \boldsymbol{Z_m}\boldsymbol{\beta_m} + \boldsymbol{Z_{rc}}\boldsymbol{\beta_{rc}} + \boldsymbol{\varepsilon}_{R*C+M} \end{aligned} \tag{S1}$$

The hat matrix for this model is denoted as $\boldsymbol{H}_{R*C+M}$, which can be computed from the QR decomposition of $\boldsymbol{Z}_{R*C+M}$ and Equation (10). Then we can compute the RSS of the model

$$\begin{aligned} RSS(R, C, M, RC) &= \boldsymbol{X}^T(\boldsymbol{I}_{N\times N} - \boldsymbol{H}_{R*C+M})\boldsymbol{X} \\ &= \boldsymbol{\varepsilon}_{R*C+M}^T(\boldsymbol{I}_{N\times N} - \boldsymbol{H}_{R*C+M})\boldsymbol{\varepsilon}_{R*C+M}. \end{aligned} \tag{S2}$$

To estimate the expectation of $RSS(R, C, M, RC)$, we need to first find out the distribution of $\boldsymbol{\varepsilon}_{R*C+M}$. Comparing to the full model, the error term $\boldsymbol{\varepsilon}_{R*C+M} = \boldsymbol{Z_{rm}}\boldsymbol{\beta_{rm}} + \boldsymbol{Z_{cm}}\boldsymbol{\beta_{cm}} + \boldsymbol{\varepsilon}$. Under the assumption of the three-way mixed effects ANOVA model, $\boldsymbol{\beta_{rm}} \sim \boldsymbol{N}(\boldsymbol{0}, \sigma_{rm}^2\boldsymbol{I}_{IJ\times IJ})$, $\boldsymbol{\beta_{cm}} \sim \boldsymbol{N}(\boldsymbol{0}, \sigma_{cm}^2\boldsymbol{I}_{IK\times IK})$, and $\boldsymbol{\varepsilon} \sim \boldsymbol{N}(\boldsymbol{0}, \sigma_\varepsilon^2\boldsymbol{I}_{N\times N})$ are independent from each other. Thus,

$$E[\boldsymbol{\varepsilon}_{R*C+M}\boldsymbol{\varepsilon}_{R*C+M}^T] = \sigma_{rm}^2\boldsymbol{Z_{rm}}\boldsymbol{Z_{rm}^T} + \sigma_{cm}^2\boldsymbol{Z_{cm}}\boldsymbol{Z_{cm}^T} + \sigma_\varepsilon^2\boldsymbol{I} \tag{S320}$$

Therefore, the expected RSS for model R, C, M, RC is

$$\begin{aligned} E[RSS(R, C, M, RC)] &= E[\boldsymbol{\varepsilon}_{R*C+M}^T(\boldsymbol{I}_{N\times N} - \boldsymbol{H}_{R*C+M})\boldsymbol{\varepsilon}_{R*C+M}] \\ &= tr\left((\boldsymbol{I}_{N\times N} - \boldsymbol{H}_{R*C+M})E[\boldsymbol{\varepsilon}_{R*C+M}\boldsymbol{\varepsilon}_{R*C+M}^T]\right) \\ &= tr\big((\boldsymbol{I}_{N\times N} - \boldsymbol{H}_{R*C+M})\boldsymbol{Z_{rm}}\boldsymbol{Z_{rm}^T}\big)\sigma_{rm}^2 \\ &+ tr\big((\boldsymbol{I}_{N\times N} - \boldsymbol{H}_{R*C+M})\boldsymbol{Z_{cm}}\boldsymbol{Z_{cm}^T}\big)\sigma_{cm}^2 + \end{aligned} \tag{S4}$$

$$tr(\boldsymbol{I}_{N\times N} - \boldsymbol{H}_{R*C+M})\sigma_\varepsilon^2$$

The RSS and its expectation for the other 15 models can be derived in a similar way.

**B. Additional Results from Simulation Studies**

Figure S1 shows the relative bias and CV for the variance estimates of BRBM differences as a function of the case variability for different levels of reader variability. These results correspond to the fourth set of parameters discussed in section 2.4 of the manuscript. As assumed in the simulation model described in Equation (11), the variances of readers and cases are $\frac{\beta_R}{\alpha_R - 1}$ and $\sigma_C^2$, respectively. Thus, the reader variability decreases as we increase $\alpha_R$, while the case variability increases as we increase $\sigma_C^2$. From Equation (15), we know that both the reader and case variability contribute to the variance of BRBM differences $V_{BR}^{12}$. We selected the parameters so that $\frac{2\beta_{\tau R}}{\alpha_{\tau R} - 1} + \frac{2\beta_R}{\alpha_R - 1} = \frac{4}{\alpha_R - 1}$ and $2\sigma_C^2$ take the values 0.2, 0.4, 0.8, 4/3, 2, indicating different levels of reader and case variability.

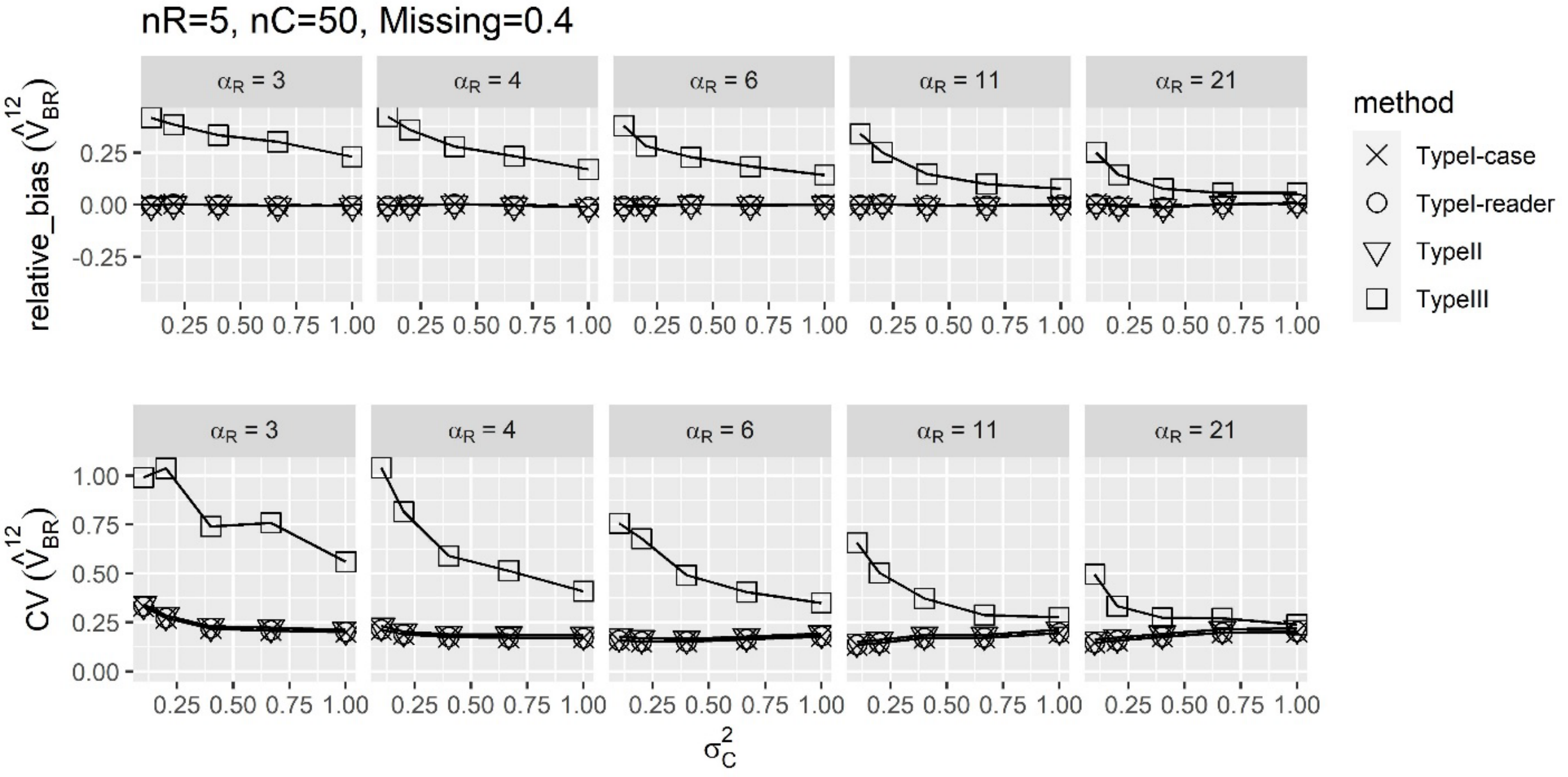

Figure S1 Relative bias and CV of variance estimates for BRBM differences as a function of the case variability for different levels of reader variability

From the first row of Figure S1, we observe that the relative bias of Type III SS methods decreases as the case variability increases and the reader variability decreases. This might be caused by there being less bias in estimating the variance component for the reader effect when the reader variability is small or relatively small compared to the case variability. The CVs for the Type III SS method always drop as the case variability increases for different level of reader variability. This tells that the uncertainty of the Type III SS method is also determined by how large the reader variability comparing to the case variability. But this does not happen for Type I or II SS method. When the case variability contributes more to $V_{BR}^{12}$ than the reader variability (last two columns of Figure S1), the CVs for the Type I or II SS method increase as the case variability increase.

Figure S2, Figure S3, Figure S4 and Figure S5 demonstrate the relative bias and CV for the variance estimates of WRBM differences for the same four sets of parameters as described in the manuscript. As shown in Table 4 of the manuscript, the Type II and Type III SS combine in the same way for the variance components and the variance of WRBM differences. Therefore, we only compare the results for three estimation methods: two Type I SS methods with different orders of effects and Type II/III SS method. The relative bias and CV for all the methods in the different parameter settings are close to each other. Due to the low resolution, the small difference might be hard to see in the figures. The relative bias observed is distributed tightly around 0 and the absolute value of the relative bias is less than 1% for most of the cases. This appears to be an artifact of the finite simulation. The CV, which explains the uncertainty of the variance estimation, decreases as the sample size increases and increases as the percentage of missing value increases.

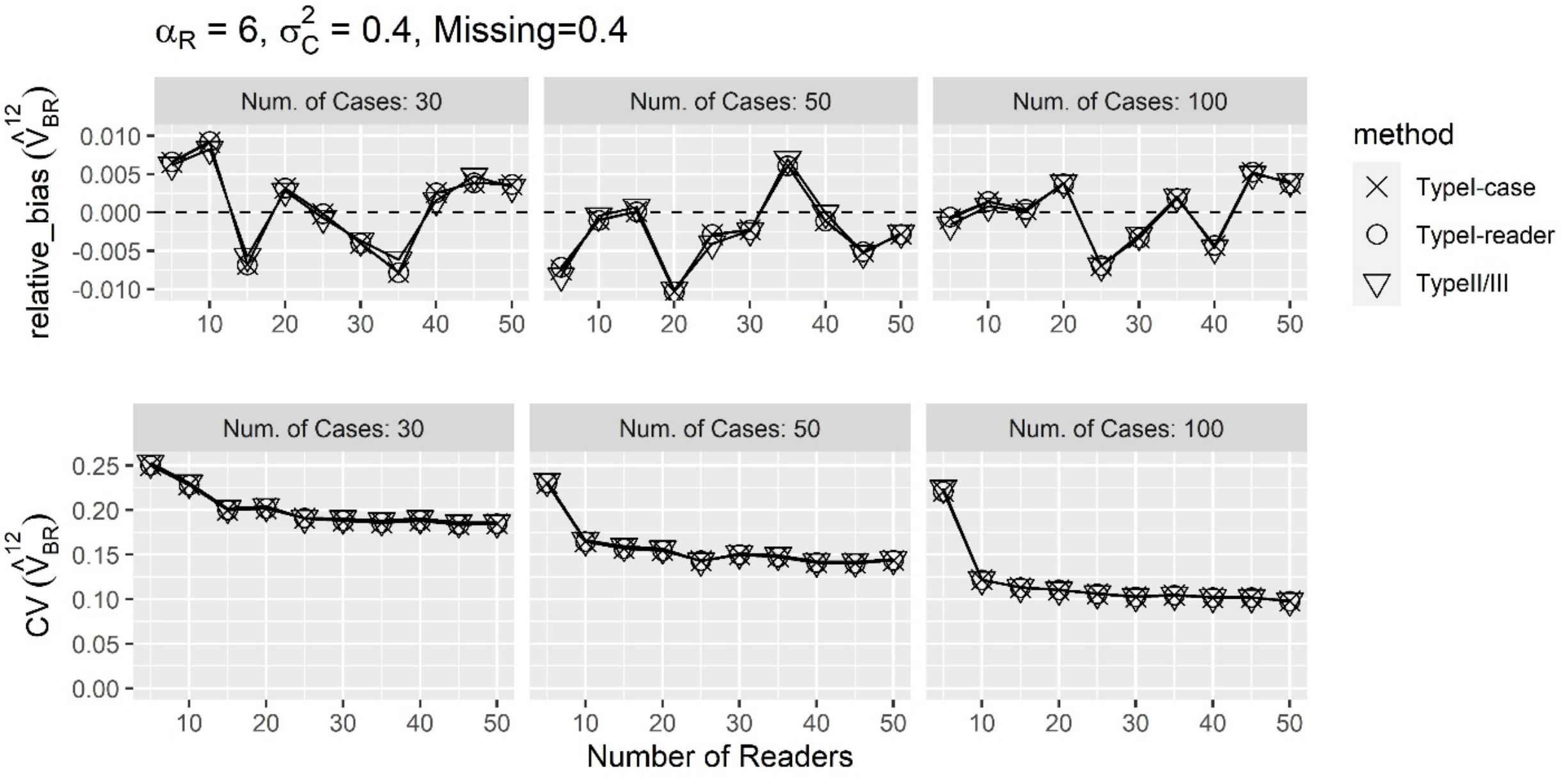

Figure S2 Relative bias and CV of variance estimates for WRBM differences changing as a function of the number of readers for different numbers of cases

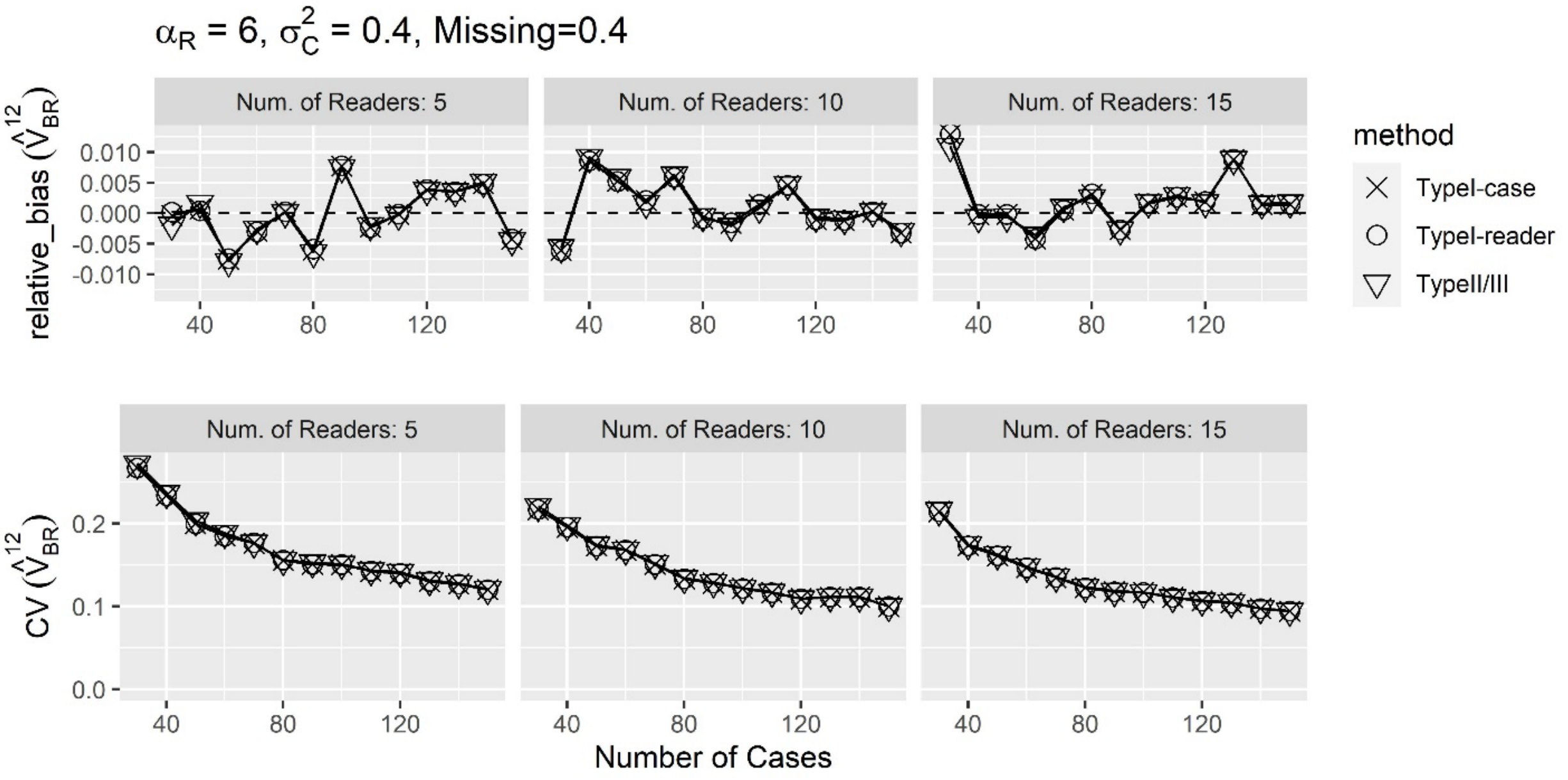

Figure S3 Relative bias and CV of variance estimates for WRBM differences as a function of the number of cases for different numbers of readers

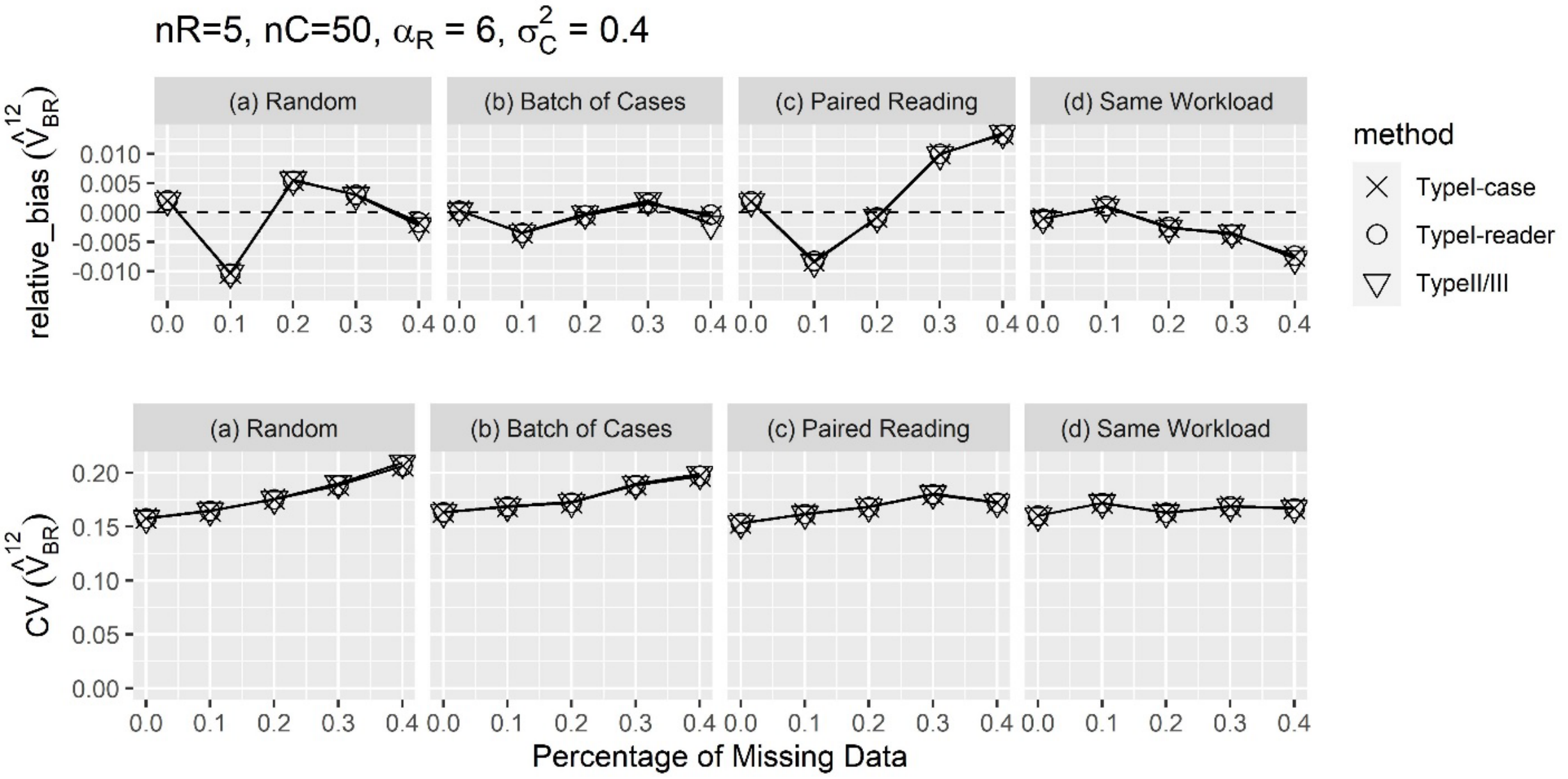

Figure S4 Relative bias and CV of variance estimates for WRBM differences as a function of the amount of missing data for different study designs

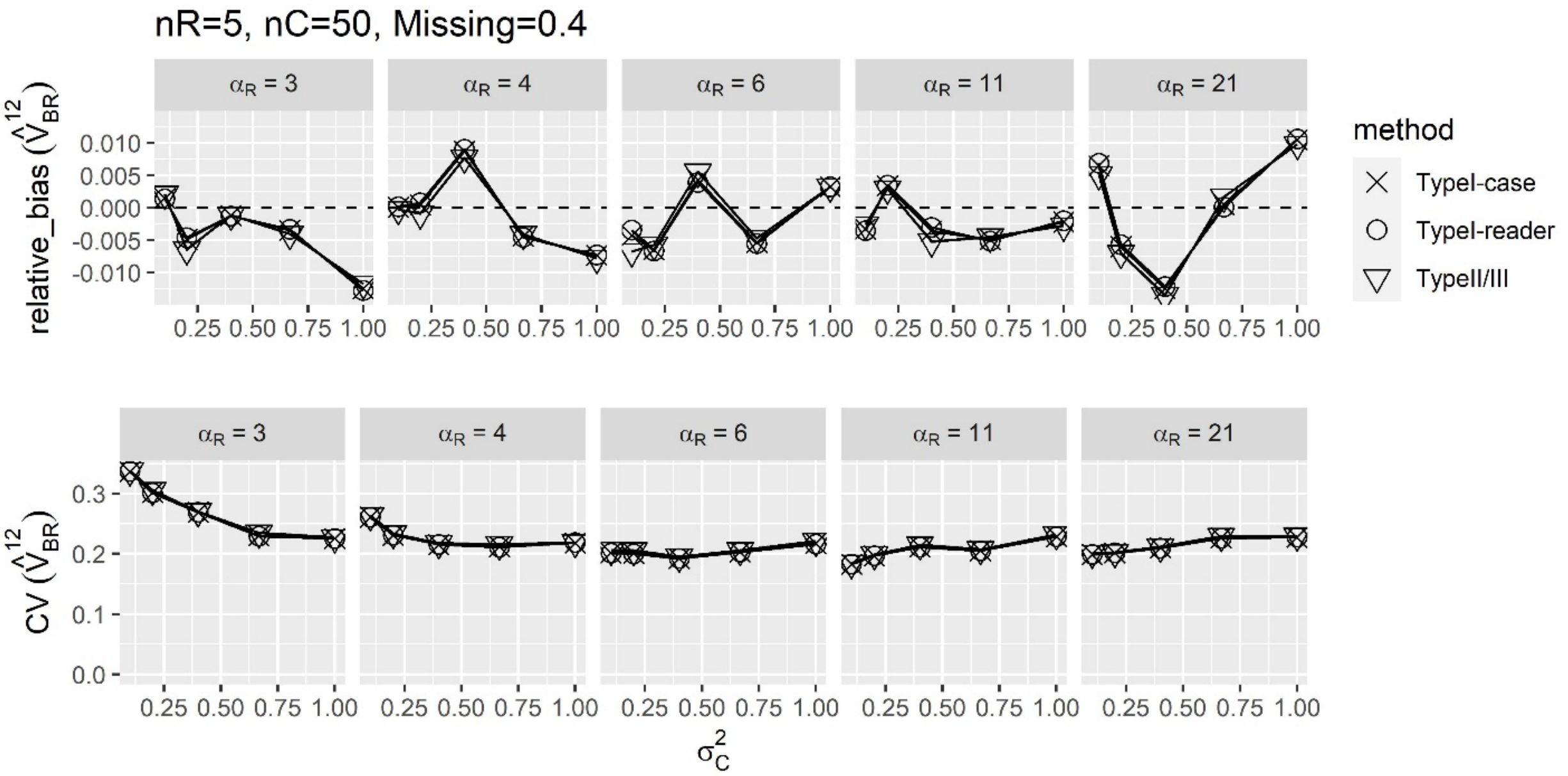

Figure S5 Relative bias and CV of variance estimates for WRBM differences as a function of the amount of case variability for different levels of reader variability

**C. Using both ANOVA models to estimate the variance of BRWM**

The BRWM limits of agreement can be estimated in two ways. One way is to apply the two-way random effect ANOVA model to the single platform/reading modality data. The variance components can be estimated as the same as those in the two-way ANOVA model for WRBM differences. Then the variance of BRWM differences can be estimated as

$$\hat{V}_{BR}^{ii} = 2\hat{\sigma}_{2w,R}^2 + 2\hat{\sigma}_{2w,\varepsilon}^2. \quad \text{(S5)}$$

The other way is to duplicate the data so that the two modalities share the same reading scores. Then apply the three-way ANOVA model to estimate the variance of BRWM difference

$$\hat{V}_{BR}^{ii} = 2\hat{\sigma}_R^2 + 2\hat{\sigma}_{RC}^2 + 2\hat{\sigma}_{mR}^2 + 2\hat{\sigma}_\varepsilon^2. \quad \text{(S6)}$$

When the scores for the two modalities are the same, the SS for modality, the interaction terms including modality, and the error will be 0 within numerical precision; in other words, $\hat{\sigma}_{mR}^2 = \hat{\sigma}_{mC}^2 = \hat{\sigma}_\varepsilon^2 = 0$. The other estimates for the variance components in the three-way ANOVA are equivalent to those in the two-way ANOVA: $\hat{\sigma}_R^2 = \hat{\sigma}_{2w,R}^2$, $\hat{\sigma}_C^2 = \hat{\sigma}_{2w,C}^2$, $\hat{\sigma}_{RC}^2 = \hat{\sigma}_{2w,\varepsilon}^2$ for Type I and II SS. For Type III SS, the estimates of the variance components are the same as Type II SS for the two-way ANOVA, but different for the three-way ANOVA, as there is no interaction term for the data with no replicates in the two-way ANOVA model. In considering the computational cost, we use the two-way ANOVA method to estimate variance of BRWM differences in our real data example as it needs to solve fewer linear regressions with smaller design matrices compared to the three-way ANOVA method.